\documentclass[aip,apl,reprint, floatfix]{revtex4-1}

\usepackage{amsmath}
\usepackage{graphicx}
\usepackage[caption=false]{subfig}

\usepackage[version=3]{mhchem}
\usepackage[]{siunitx}

\begin{document}

\title{Negative Differential Resistance and Steep Switching in Chevron Graphene Nanoribbon Field Effect Transistors }

\author{Samuel Smith, Juan-Pablo Llin\'as, Jeffrey Bokor, Sayeef Salahuddin}
\affiliation{University of California, Berkeley, Department of Electrical Engineering and Computer Sciences, Berkeley, CA 94720}

\begin{abstract}
Ballistic quantum transport calculations based on the non-equilbrium Green's function formalism show that field-effect transistor devices made from chevron-type graphene nanoribbons (CGNRs) could exhibit negative differential resistance with peak-to-valley ratios in excess of 4800 at room temperature as well as steep-slope switching with 6 mV/decade subtheshold swing over five orders of magnitude and ON-currents of \SI{88}{\micro\ampere\per\micro\meter}. This is enabled by the superlattice-like structure of these ribbons that have large periodic unit cells with regions of different effective bandgap, resulting in minibands and gaps in the density of states above the conduction band edge. The CGNR ribbon used in our proposed device has been previously fabricated with bottom-up chemical synthesis techniques and could be incorporated into an experimentally-realizable structure.
\end{abstract}


\maketitle

\newlength\figureheight
\newlength\figurewidth
\setlength\figureheight{5 cm}
\setlength\figurewidth{6.0 cm}

In 1970, L.Esaki and R. Tsu predicted \cite{esaki1970superlattice} that in an appropriately made superlattice, it should be possible to obtain very narrow width bands, which could then lead to negative differential resistance. The remarkable property of these superlattices is in the fact that, unlike the Esaki diodes, this negative differential resistance does not need any tunneling, rather it comes from the direct conduction of electrons. Nonetheless, significant difficulty in synthesizing atomically precise, eptiaxial heterostructures has made it very challenging to realize such superlattice structures \cite{esaki1974new, choi1987periodic, ismail1989negative, bernstein1987negative, grahn1991electric, kastrup1994multistability, warren1985surface}. Much work has been done on modeling graphene nanoribbon heterostructures and superlattices which could exhibit NDR \cite{ferreira2011low, li2015low, teong2009shape, sevinccli2008superlattice, sharifi2016negative, saha2015double, girao2012electronic}. Other work has been done on steep slope devices based on GNR and CNT heterojunctions\cite{kim2014computational,yoon2010barrier}.  Gnani \textit{et al.} showed how superlattices could be used in a III-V nanowire FET to achieve steep slope behavior by using the superlattice gap to filter high energy electrons in the OFF state \cite{gnani2011performance}. Here, we show that the recently synthesized chevron nanoribbons\cite{cai2010atomically} provides a natural, monolithic material system where narrow-width energy bands and negative differential resistance (NDR) can be achieved. Our atomistic calculations predict that the NDR behavior should manifest at room temperature along with sub-thermal steepness ($<$60 mV/decade at room temperature). Such NDR behavior could lead to completely novel devices for next generation electronics. 

Unlike a graphene sheet, a narrow strip etched out of graphene, often called a graphene nanoribbon (GNR), can provide a sizeable bandgap. As a result, GNRs could lead to devices with good ON/OFF ratio at the nanoscale. However, a number of studies have also shown the deleterious effect of edge roughness on the device performance \cite{fiori2007simulation, yoon2007effect}. Recent breakthroughs in bottom-up chemical synthesis can produce GNRs with atomistically pristine edge states and  overcome this shortcoming\cite{cai2010atomically}. In fact, a recent experimental work demonstrated working transistors with 9- and 13-AGNRs made with these techniques\cite{llinas2016short}. The methods used to synthesize these ribbons can also be used to generate complex periodic structures beyond simply armchair and zigzag nanoribbons\cite{cai2014graphene, chen2015molecular}. In this work, we will consider one of those structures, the chevron graphene nanoribbon (CGNR).

\begin{figure}[b]
	\centering
	\includegraphics[width =\columnwidth]{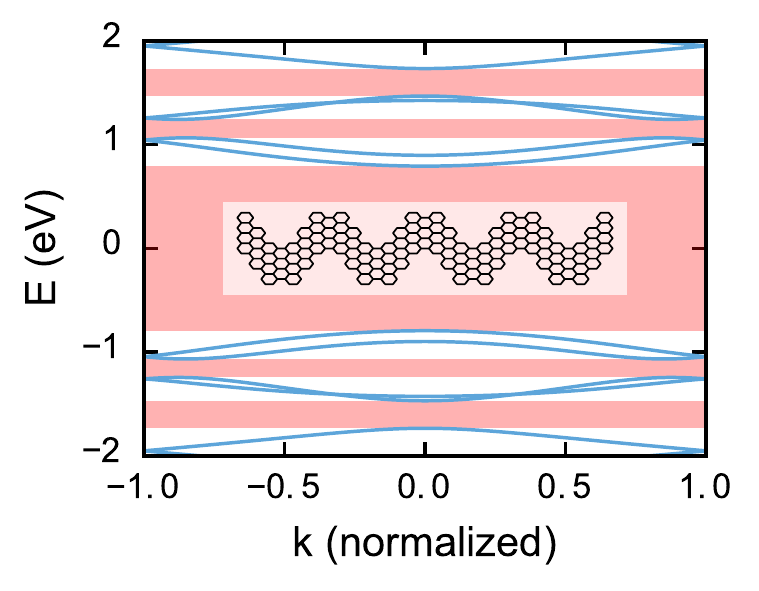}
	\caption{Band structure of a chevron graphene nanoribbon based on a $p_z$ orbital basis set. The width (and thus quantum confinement) varies across the unit cell, giving a superlattice-like band structure. Forbidden energies are highlighted in red. The bandgap of the ribbon is 1.59 eV, the first conduction band has a bandwidth of 0.272 eV, and the first gap between minibands is 0.178 eV. Inset: Molecular structure of the chevron nanoribbon.}
	\label{fig:bandsLattice}
\end{figure}

Fig. \ref{fig:bandsLattice} shows both the atomic structure of the 6-9 CGNR originally fabricated by Cai \textit{et al.} and the electronic structure calculated through a $p_z$ orbital-based tight-binding method\cite{cai2010atomically}. A key feature of the band structure is the presence of minibands with regions of forbidden energy above the conduction band edge, such as those seen in superlattices of III-V semiconductors. Analogous to III-V superlattices, the CGNR contains regions of different effective bandgap.  When we look at the CGNR in Fig. \ref{fig:bandsLattice}, we see that its narrowest segment is 6 carbon atoms across and its widest segment is 9 carbon atoms across, with both segments having armchair-type edges. Using a pz-basis set (GW\cite{yang2007quasiparticle}), the bandgap, $E_g$, of a 6-AGNR is 1.33 eV (2.7 eV) and the bandgap of a 9-AGNR is 0.95 eV (2.0 eV). However, given the very short length scale over which the width changes in our structure ($\sim \SI{1}{\nano\meter}$), one would not expect the system to behave as though the local effective potential oscillates between the bulk values of $E_g$ for the isolated AGNRs. In fact, our chevron structure has an overall bandgap of 1.59 eV. This value is consistent with the \SI{1.62}{\electronvolt} bandgap from LDA DFT calculations, but significantly smaller than the \SI{3.74}{\electronvolt} value from calculations incorporating the GW correction\cite{wang2012quasiparticle}. Both LDA and GW calculations show the presence of minibands and gaps above the conduction band edge\cite{wang2012quasiparticle}.

\begin{figure}[t]
	\centering
	\includegraphics[width =\columnwidth]{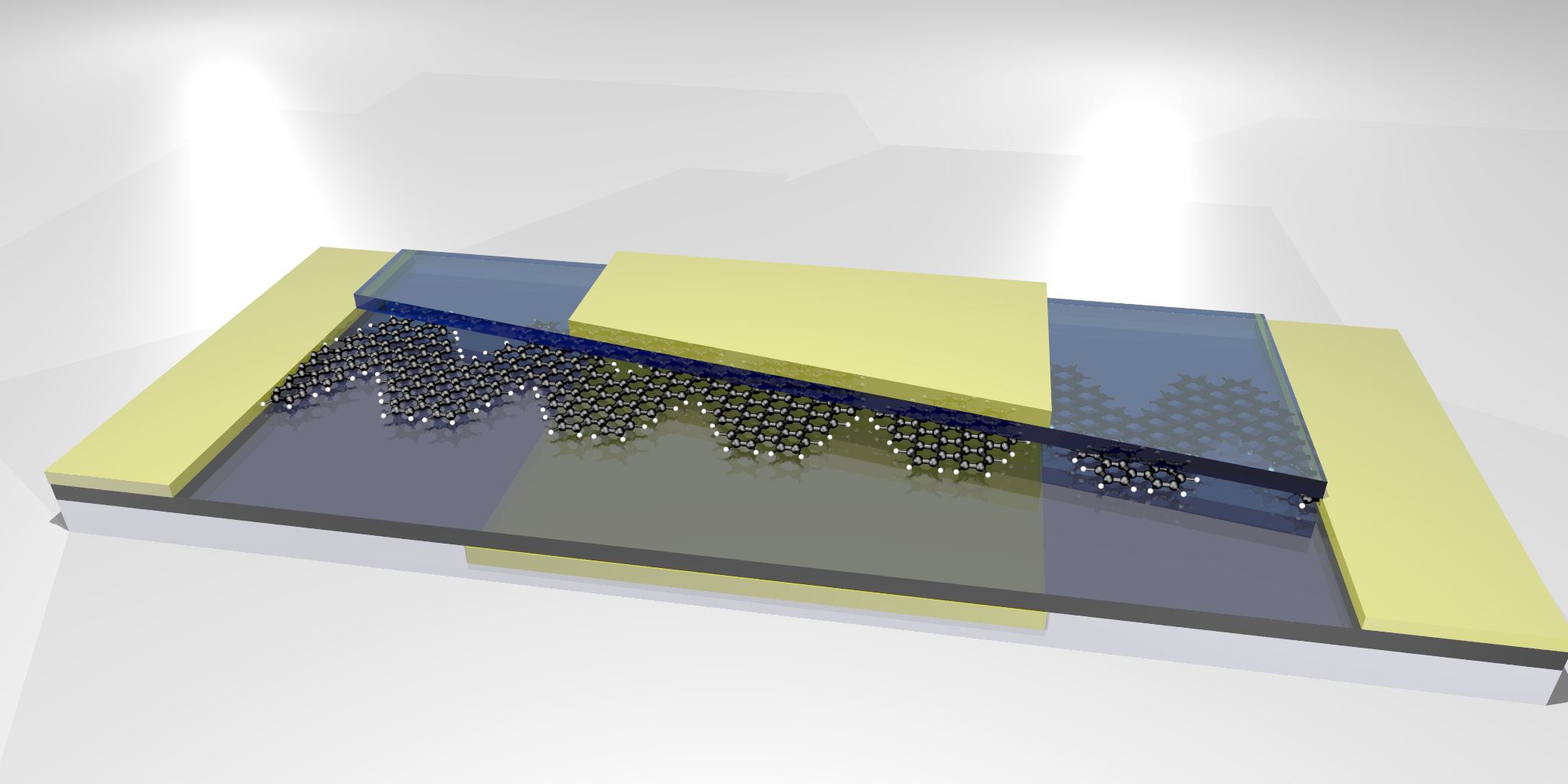}\\
    \vspace{6pt}
    \includegraphics[width =\columnwidth]{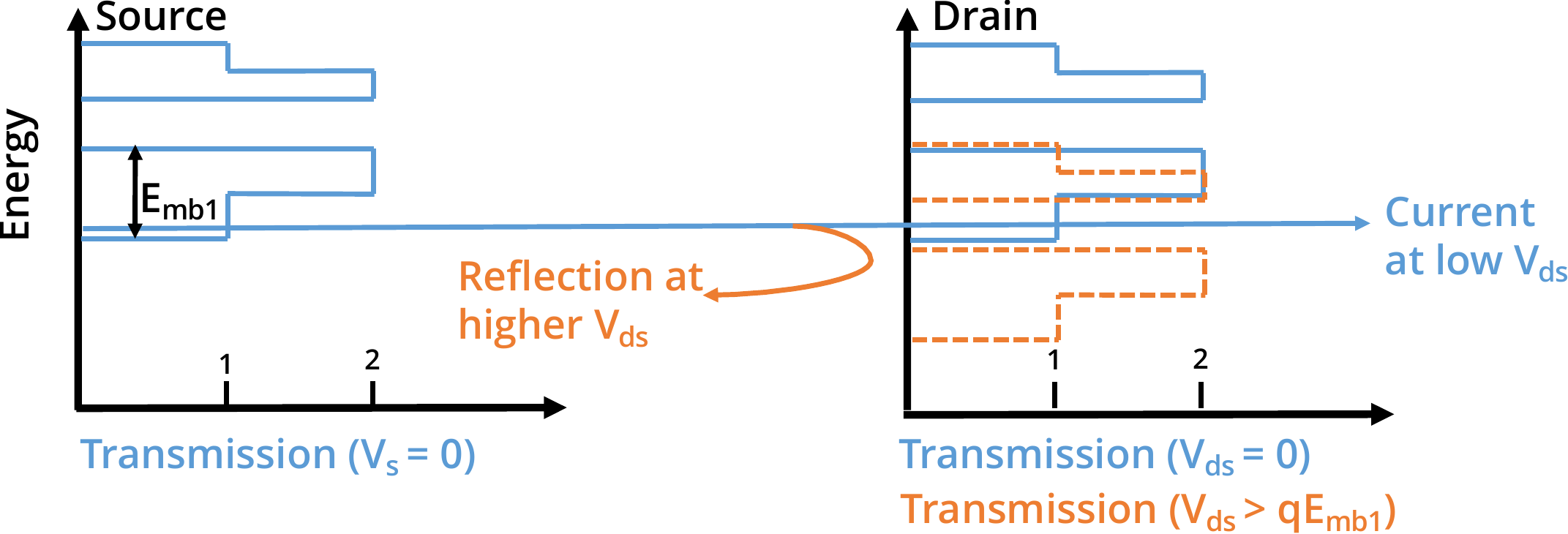}
	\caption{Artistic rendering of double-gate CGNR on insulator SLFET. Parts of the top gate and oxide region have been cut away so that the channel is visible. When a gate voltage is applied to turn the device ON, current conduction occurs at low values of $V_{ds}$ where the first miniband at the source is aligned with the first miniband at the drain. As the drain voltage is increased beyond $qE_{mb1}$, the bandwidth of the first miniband, transmission from source to drain is cut off and the device exhibits negative differential resistance.}
	\label{fig:DeviceExplanation}
\end{figure}

The structure of the simulated device is shown in in Fig. \ref{fig:DeviceExplanation}. Like a typical MOSFET, our superlattice field-effect transistor (SLFET) can be turned ON and OFF with a gate voltage at low drain biases. Operation differs from a MOSFET in two key ways. The first is that the device shows NDR with respect to the drain voltage. At some value of $V_{ds}$ determined by the width of the first miniband, $I_d$ decreases substantially when the conduction band at the source becomes aligned with the superlattice gap at the drain. At higher drain bias, current increases again when the conduction band at the source is aligned with the second miniband at the drain. The second feature of the SLFET is that the superlattice gap at the drain filters out higher energy electrons from the first miniband at the source when a source-drain bias is applied. This cuts off the higher energy portion of the thermionic tail at the source, which would contribute to leakage current in a traditional MOSFET. This filtering does not, however, affect the low-energy electrons, which carry most of the ON state current as they are in the window where the first minibands at the source and drain overlap. Transport in an SLFET is entirely intraband like a MOSFET, whereas a TFET relies on band-to-band tunneling. This could possibly allow higher ON current than a TFET.

The CGNR used in our simulation has a width of 1.9 nm. The simulation domain is approximately 70 nm long, and the gate has a length of 15 nm. The source and drain are doped with $N_{D} = 1.0\times 10^{13} \;\mathrm{cm^{-2}}$ donors. An effective oxide thickness of \SI{1.0}{\nano\meter} is used for both the top and bottom gates. The gate contacts are extended \SI{30}{\nano \meter} in the direction perpendicular the channel to capture the effects of fringing gate fields. While our simulation uses an effective doping density to align the source and drain Fermi levels to the CGNR conduction band, the same effect could be achieved in an experiment through electrostatics alone.

Simulations are performed using the non-equilibrium Green’s function (NEGF) formalism\cite{datta2005quantum}. A simple $p_z$-basis is used for the Hamiltonian for the chevron graphene nanoribbon with the hopping parameter set to $t_0 = \SI{2.7}{\electronvolt}$. Charge and current are calculated with the recursive Green’s function algorithm\cite{lake1997single}, and contact self-energy is computed with the Sancho-Rubio iteration scheme\cite{sancho1985highly}. 

The NEGF equations are solved self-consistently with the Poisson equation for electrostatics in three dimensions. Our simulator solves the nonlinear Poisson equation using the predictor-corrector scheme described by Trellakis \textit{et al}. using a semiclassical approximation for the charge density\cite{trellakis1997iteration}. The geometry of the system is modeled using a tetrahedral finite element mesh generated with the SALOME package\cite{ribes2007salome}. The solution of the final sparse matrix form of the Poisson equation discretized the with finite element method is performed using the conjugate gradient solver from the Eigen library\cite{eigenweb}.

\begin{figure*}[t]
	\subfloat[$V_{gs} = \SI{0.55}{\volt}, V_{ds}=\SI{0.10}{\volt}$ \label{sfig:dos_idvg_25_1}]{%
		\includegraphics[width=.3\linewidth]{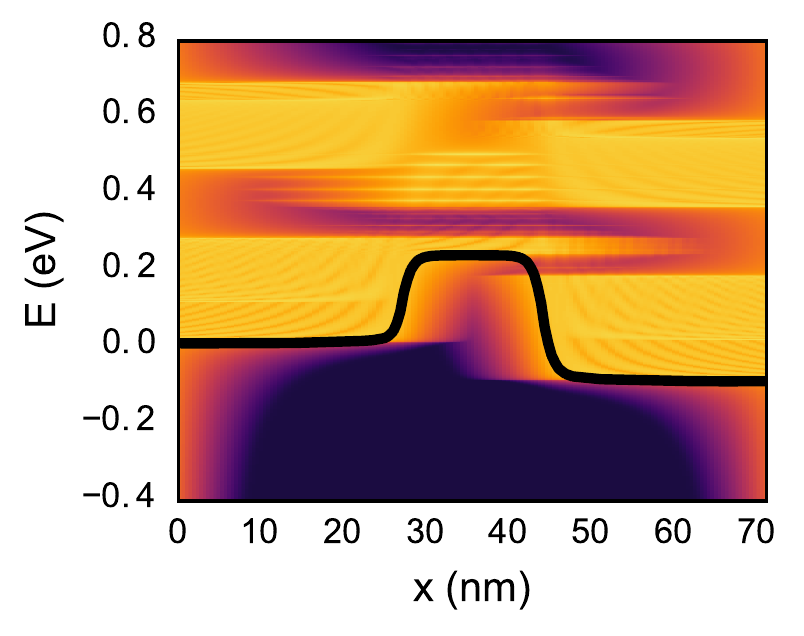}%
	}\hfill
	\subfloat[$V_{gs} = \SI{0.70}{\volt}, V_{ds}=\SI{0.10}{\volt}$ \label{sfig:dos_idvd_2_10}]{%
		\includegraphics[width=.3\linewidth]{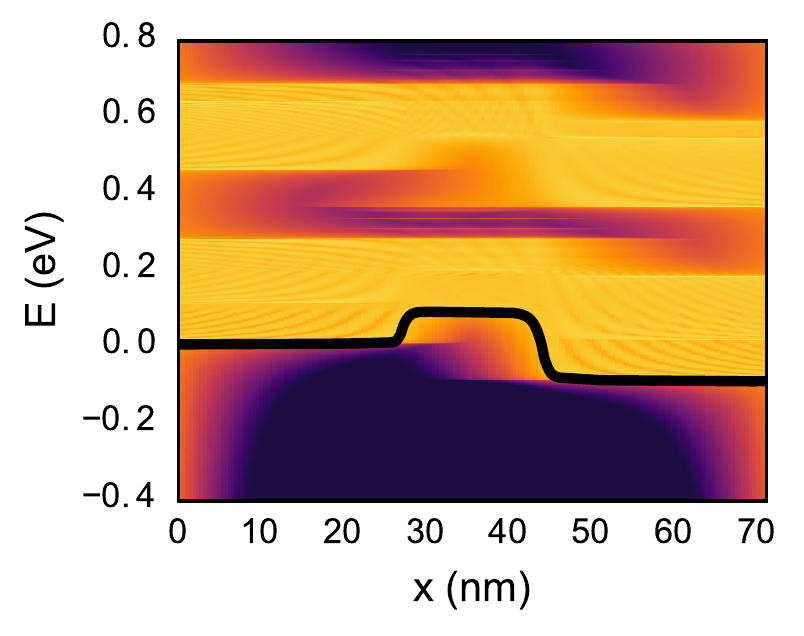}%
	}\hfill
	\subfloat[$V_{gs} = \SI{0.70}{\volt}, V_{ds} =\SI{0.3}{\volt}$\label{sfig:dos_idvd_2_30}]{%
		\includegraphics[width=.3\linewidth]{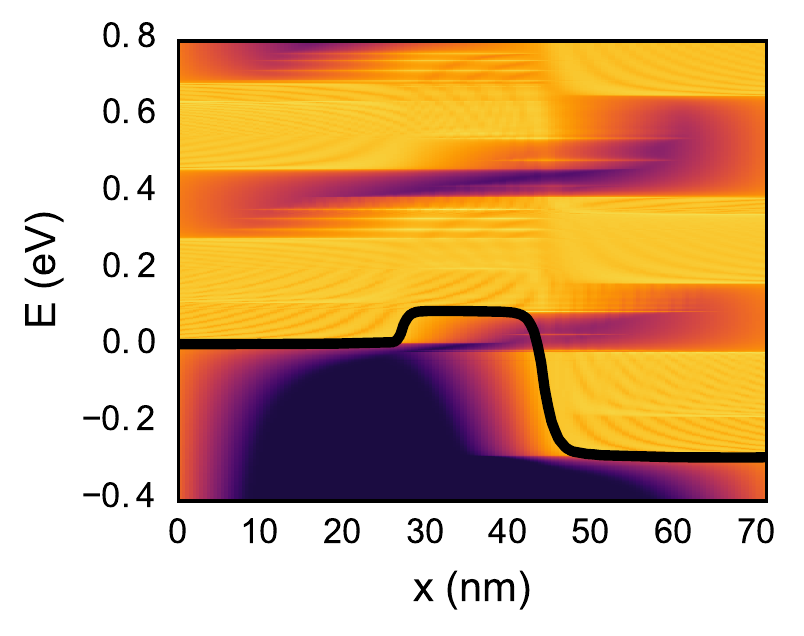}%
	}\hfill
	\caption{Local density of states for several different biasing conditions. Fig. (a) shows the OFF state, in which leakage current is substantially reduced because the superlattice gap in the drain region filters higher energy electrons, which could otherwise travel over the source-side barrier. Fig. (b) shows the ON state, in which current is primarily carried by lower energy carriers, which are not blocked through the density of states filtering at the drain. Fig. (c) shows the ON state for a higher value of $V_{ds}$. Significant ballistic transport from source to drain is no longer possible when the drain voltage is greater than the width of the first miniband minus the height of the source-side barrier. The colormap is based on a logarithmic scale.}
	\label{fig:ldos}
\end{figure*}

The local density of states for the CGNR MOSFET is shown in Fig. \ref{fig:ldos} for several biasing conditions. Fig. \ref{sfig:dos_idvd_2_10} shows the case for peak current for the device when a large enough drain bias has been applied to generate enough splitting between the source and drain Fermi levels to allow significant current to flow, but not a high enough bias to move the the first miniband outside of the current conduction window. For higher bias as in Fig. \ref{sfig:dos_idvd_2_30}, intraband conduction from the first miniband is completely cut off. As the drain bias is further increased, current can only flow due to a band-to-band tunneling from the first miniband at the source to the second miniband at the drain. Note that, due to the minibands, there will be regions of operation for both gate and drain voltages where current flow is abruptly turned on or off, as the overlap between source minibands and drain minibands is modified. This leads to a steep subthreshold swing ($<$ 60 mV/decade at room temperature) in the $I_d-V_g$ characteristic and a negative differential resistance in the $I_d-V_d$ characteristic. 

\begin{figure}[pt]
	\centering
	\includegraphics[width =\columnwidth]{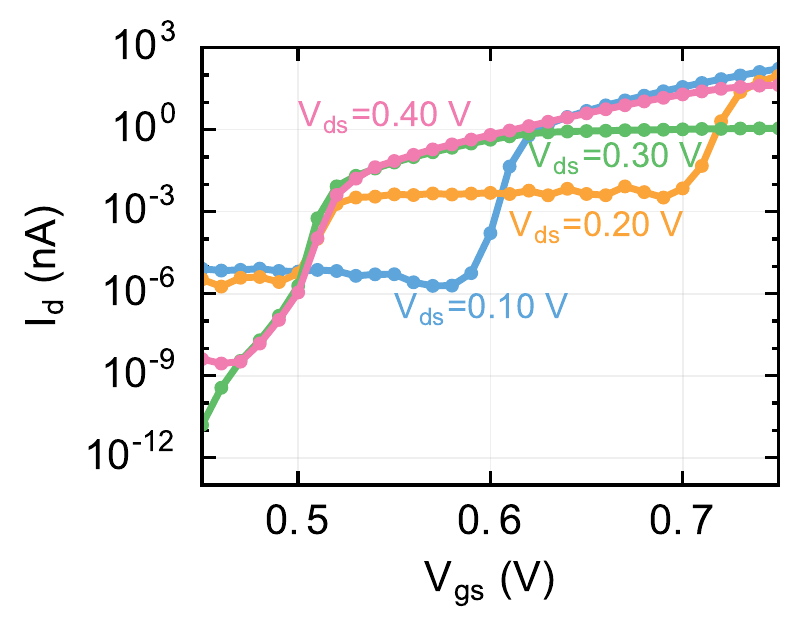}
	\caption{$I_d-V_{gs}$ plot for different values of $V_{ds}$. Steep-slope behavior is observed with a subtheshold swing of around 6 mV/decade over five orders of magnitude around $V_{gs} = \SI{0.6}{\volt}$ for the case when $V_{ds} = \SI{0.1}{\volt}$. Negative differential resistance is evident in that the peak value of $I_d$ is lower for higher values of $V_{ds}$.}
	\label{fig:IdVg}
\end{figure}

We shall first discuss the current vs gate voltage ($I_d-V_g$) characteristics. Fig. \ref{fig:IdVg} shows $I_d$ vs. $V_{gs}$ for several values of $V_{ds}$. While steep slope behavior is exhibited at some point for all values of $V_{ds}$, the highest ON current is obverved for $V_{ds} = \SI{0.1}{\volt}$. At this drain bias, an ON current of \SI{168}{\nano\ampere} (\SI{88}{\micro\ampere\per\micro\meter}) is achieved at a gate bias of $V_{gs} = \SI{0.75}{\volt}$. In the steep slope region of this curve, the subthreshold swing is 6 mV/decade when averaged over five orders of magnitude of $I_d$. With gate work function engineering and additional device optimization, it should be possible to achieve reasonable ON current with a low supply voltage in devices of this type. The origin of the steep-slope behavior can be understood from Fig. \ref{fig:ldos}. In the OFF state shown in Fig. \ref{sfig:dos_idvg_25_1}, the superlattice gap at the drain prevents leakage current from flowing over the source-side injection barrier. The states near the top of the barrier are seen to decay rapidly in the drain region. Fig. \ref{sfig:dos_idvd_2_10} shows the ON state, in which low-energy electrons, which make up virtually all of the ON current, can flow unimpeded from source to drain.

\begin{figure}[t]
	\centering
	\includegraphics[width =\columnwidth]{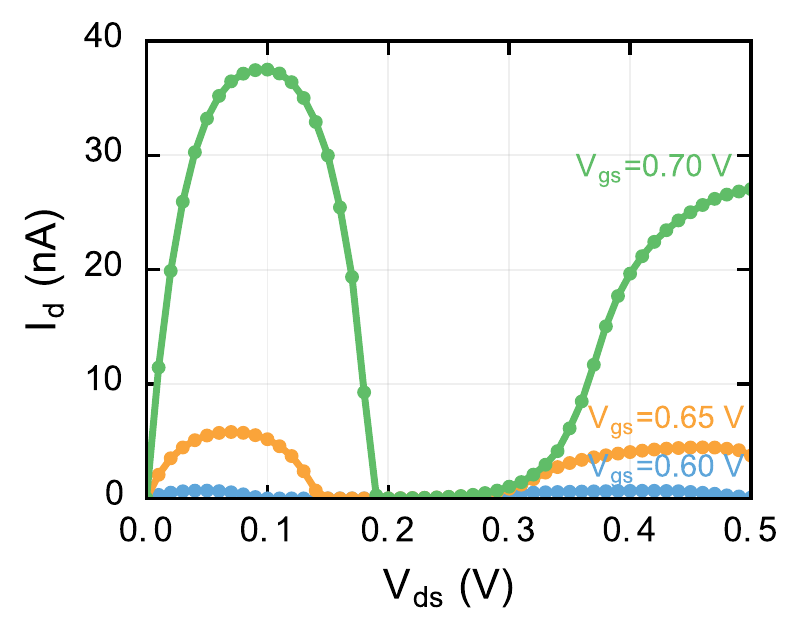}
	\caption{$I_d-V_{ds}$ plot for different values of $V_{gs}$. For the case, when $V_{gs} = \SI{0.7}{\volt}$, a peak-to-valley ratio of $4.88\times 10^3$ is achieved.}
	\label{fig:IdVd}
\end{figure}

The $I_d-V_{ds}$ curves from the results of our simulation are shown in Fig. \ref{fig:IdVd}. Considering the case when $V_{gs} = \SI{0.7}{\volt}$, we see an increase in current up to $V_{ds} = \SI{0.10}{\volt}$. As the drain bias is further increased, we see a decrease in current as the drain miniband goes out of alignment with the source miniband. The current beigins to pick up again as the second miniband at the drain starts to come in alignment with the source miniband again. The peak-to-valley ratio (PVR) at this gate voltage is $4.88\times 10^3$. At $V_{gs} = \SI{0.60}{\volt}$, the calculated PVR is $1.71\times 10^8$. Note, however, that this value is expected to much smaller in a practical device due to the electron-phonon scattering mechanisms\cite{lake1997single} that were not taken into account in our ballistic simulations. 

In summary, we have shown that chevron graphene nanoribbon devices can exhibit both steep-slope subthreshold behavior and negative differential resistance. Both properties are the result of the superlattice-like electronic structure of the ribbon. CGNR SLFETs could be promising for a number of applications ranging from low-power logic transistors to high speed oscillators. A major obstacle to building a real device is making contacts with appropriate Schottky barrier heights to be able to match the band alignment conditions achieved in this work through a simple doping model. The performance of a real device would also likely be impacted by scattering mechanisms we have not considered here, though the ability to synthesize ribbons with virtually no defects may minimize these effects. Additional optimization will also likely be necessary to make a functioning device. DFT+GW calculations predict a much higher bandgap for the CGNR in vacuum than the tight-binding model used in this work. While surface screening may reduce the bangap somewhat, a wider ribbon with a narrower bandgap may be required. Co-optimization of the bandgap with the bandwidths of the minibands and the gaps between minibands is also a necessary topic for future work.

\section*{Acknowledgements}
This work was supported by NSF CAREER grant CISE-1149804. Work by JPL and JB was supported in part by the Office of Naval Research BRC program under Grant N00014-16-1-2229. The authors would like to acknowledge useful discussions with Felix Fischer and Michael Crommie.

\bibliography{references}

\end{document}